\documentclass[aps,prl,twocolumn,superscriptaddress]{revtex4}
\usepackage{amsmath}
\usepackage{graphicx}
\usepackage{epstopdf}
\usepackage{dcolumn}
\usepackage{bm}
\usepackage{amsfonts}
\usepackage{color}
\makeatletter
\def\captionof#1#2{{\def\@captype{#1}#2}}
\makeatother

\begin{document}


\title{Fluctuating hydrodynamics for a discrete Gross-Pitaevskii equation: mapping to Kardar-Parisi-Zhang universality class}
\author{Manas Kulkarni}
\affiliation{Department of Physics, New York City College of
Technology, The City University of New York, Brooklyn, NY 11201, USA}
\author{David A. Huse}
\affiliation{Department of Physics, Princeton University, Princeton, NJ 08544, USA}
\author{Herbert Spohn}
\affiliation{Institute for Advanced Studies, Einstein Drive, Princeton NJ 08540, USA}
\affiliation{Zentrum Mathematik and Physik Department, TU M\"unchen,\\
 Boltzmannstrasse 3, D-85747 Garching, Germany}
\begin{abstract}
We show that several aspects of the low-temperature hydrodynamics of a discrete Gross-Pitaevskii equation (GPE) can be understood by mapping it to a nonlinear version of fluctuating hydrodynamics.  This is achieved by first writing the GPE in a hydrodynamic form of a continuity and an Euler equation.  Respecting conservation laws, dissipation and noise due to the system's chaos are added, thus giving us a nonlinear stochastic field theory in general and the Kardar-Parisi-Zhang (KPZ) equation in our particular case.  This mapping to KPZ is benchmarked against exact Hamiltonian numerics on discrete GPE by investigating the non-zero temperature dynamical structure factor and its scaling form and exponent.  Given the ubiquity of the  Gross--Pitaevskii equation (a.k.a. nonlinear Schr\"{o}dinger equation), ranging from nonlinear optics to cold gases, we expect this remarkable mapping to the KPZ equation to be of paramount importance and far reaching consequences.

\end{abstract}

\maketitle

\setcounter{equation}{0}

\textit{Introduction:} Low dimensional classical and quantum systems are often very counter-intuitive and different from their higher dimensional counterparts \cite{RevModPhys.84.1253}. One such example is the width of the line-shape of the phonon peaks in the dynamical structure factor. Contrary to the expected $k^2$ behavior in higher dimensions \cite{dfb}, the power is anomalous in low dimensions. Linearized hydrodynamics, which predicts a diffusive broadening, fails in one dimension (1D). This immediately creates a need for a nonlinear hydrodynamics that could describe low-dimensional systems. Such a theory beyond the conventional Luttinger Liquid would describe the super-diffusive broadening in low dimensional systems.

An experimentally realized \cite{bg1d} system that provides a remarkable platform for probing low-dimensional fluids is the system of a 1D weakly interacting Bose gas at non-zero temperature. Using a variant of Bragg spectroscopy \cite{fab1,fab2} one could probe the dynamical structure factor of the Bose gas, thereby unraveling the nonlinear phenomenon in low-dimensional fluids. Needless to say, the underlying theory that describes \cite{gpeder} this cold atomic system, namely, the Gross--Pitaevskii equation (GPE) or the nonlinear Schr\"{o}dinger (NLS) equation is ubiquitous in areas such as optics, cold gases and mathematical physics.  Although the strictly continuum GPE is integrable, the experimental realizations break integrability in one or more ways such as, presence of a lattice or trapping potential, energy loss, escape or unwanted evaporation of particles.
Here we focus in particular on the discrete (lattice) version of GPE which is not integrable; such a discrete GPE has been realized in experiments on waveguide lattices \cite{nwl}.

The ubiquity of such equations and cutting edge technologies available to probe statistical properties 
of such systems, enhances an urgent need for writing a stochastic nonlinear theory that makes transparent the role of various components that result in a complex nonlinear-driven-dissipative phenomenology.  Establishing this strong connection between GPE and stochastic nonlinear differential equations (which turns out to be a two-component KPZ equation in our case) helps in using the tools available in the literature to make far reaching predictions about the statistical mechanics of systems such as a 1D Bose gas or optical waveguides.  In the converse, one could also use such systems as an experimental test bed for KPZ phenomena, providing much needed additional experimental realizations of KPZ physics \cite{ex1,ex2,ex3}.\\
\\
In this Letter, we analyze the low-temperature hydrodynamics of the GPE, which is known to be a valid description for systems such as a 1D weakly interacting
Bose gas or optical waveguides.
We present a discrete GPE that governs the dynamics of such complex wave fields (which are atomic fields in the case of cold atoms or optical fields in waveguides).
We write down continuity-like and Euler-like equation for the macroscopic density and velocity fields and derive the nonlinear
fluctuating hydrodynamics. The coefficients of the resulting nonlinear fluctuating theory are expressed in terms of the underlying parameters of the system
(such as coupling strength, background density). Having established this, we present results for the dynamical structure factor $S(k,\omega)$
(i.e., fourier transform of correlation function of fields obeying nonlinear fluctuating hydrodynamic theory), namely its scaling function and the
underlying anomalous exponent.
This effective nonlinear hydrodynamic theory is finally benchmarked against exact Hamiltonian numerics, which also supports a recent remarkable conjecture that the long-wavelength dynamics of a classical 1D fluid at finite temperature is in the Kardar-Parisi-Zhang (KPZ) universality class~\cite{hvb2012}.
In addition to the confirmation of the $3/2$ exponent, we have taken a big step forward in showing agreement with the Prahofer-Spohn scaling function~\cite{ps04}. Therefore, the notoriously difficult problem of computing the dynamical structure factor (or density-density correlations) can now be connected to correlation functions of familiar stochastic differential equations.
We discuss certain points such as role of integrability, analysis of Lyapunov exponents, and future challenges and present the consequences of this mapping and its possible exploitation in understanding experiments governed by the Nonlinear Schr\"{o}dinger equation. \\
\textit{Nonlinear fluctuating hydrodynamics and GPE:}\label{sec2}
\setcounter{equation}{0}
The semi-classical Hamiltonian describing a strictly one-dimensional gas of bosons of mass $m$ and contact interaction strength $g$ is given by
\begin{eqnarray}
\label{congpe}
H=\int dx\left[\frac{\left|\partial_{x}\psi\right|^{2}}{2m}+\frac{g}{2}|\psi|^{4} \right] ~.
\end{eqnarray}
which in conjugation with Poisson brackets $\{ \psi^{*}(x),\psi(y) \}=i\delta(x-y)$ gives the time-dependent GPE,
\begin{eqnarray}
\label{tGPE}
i\partial_{t}\psi=-\frac{1}{2m}\partial_{x}^{2}\psi+g|\psi|^{2}\psi~.
\end{eqnarray}	
This is an integrable system.  But all physical realizations are not in this ideal limit: they may be in a lattice rather than with continuous translational invariance,
and they are not strictly one-dimensional, so the interaction has a nonzero range.  Here we will assume we are not in the ideal
integrable limit, so integrability is destroyed and this nonlinear classical system is chaotic at nonzero temperature.  The specific integrability breaking we consider
is the discrete GPE (equivalently, NLS) on a one-dimensional lattice (see Eq. \ref{2.2})
but the results should apply more generally.
For optical applications, $g$ is a Kerr nonlinearity and $|\psi|^{2}$ is the intensity of the light field. 

We examine the hydrodynamics of the equilibrium steady-state that this chaotic system approaches at long times.
We are interested in the hydrodynamic scaling of the density-density correlation
\begin{equation}\label{2.3}
S(x,t) = \langle |\psi(x,t)|^2|\psi(0,0)|^2\rangle - \langle |\psi(0,0)|^2\rangle^2
\end{equation}
with $\langle \cdot \rangle$ denoting the average over the statistical steady state.  $\psi(x,t)=\sqrt{\rho(x,t)}e^{i \theta(x,t)}$ defines the density $\rho(x,t)$ and the phase $\theta$.  The velocity is $v(x,t)=\frac{1}{m}\frac{\partial\theta(x,t)}{\partial x}$.  We work at low enough temperature that the rate at which phase slips occur at equilibrium is negligible, so the velocity is a conserved quantity, as is the density.
The continuity and Euler equations are 
\begin{equation}\label{2.4}
\partial_t \rho+ \partial_x(\rho v) = 0\,,\quad \partial_t v +\partial_x\bigg(\frac{v^2}{2} + \tfrac{g}{m}  \rho\bigg) = 0\,.
\end{equation}
The equilibrium state has average
density $\rho_0 = \langle |\psi(x)|^2\rangle$ and we consider the case of zero average velocity.

In the regime we are considering, Eq. (\ref{2.3}) refers to small deviations
from the average density. Hence if we linearize Eq. (\ref{2.4}), taking $\rho \rightarrow \rho_0 +\varrho$ and $v \rightarrow 0+v$,
we obtain 
\begin{equation}\label{2.5}
\partial_t \vec{u}+ \partial_x A
\vec{u}\,\,\,\quad\mbox{with}\quad
\vec{u}=\left(\begin{array}{c}\varrho\\v\end{array}\right)
,\quad A =
\begin{pmatrix}
0&\rho_0 \\ \tfrac{g}{m}& 0
\end{pmatrix}
\,.
\end{equation}
Eq. (\ref{2.5}) gives right- and left-moving sound modes with speed $c = \sqrt{g\rho_0/m}$.
One can view Eq. (\ref{2.5}) as the dynamics of a linearized Luttinger liquid whose $S(k,\omega)$ consists only of a pair of delta function peaks at $\omega=\pm c|k|$, corresponding to undamped phonons.  We need to add to this, the scattering between the phonons due to the nonlinearities.
In linear fluctuating hydrodynamics one adds damping and noise to Eq. (\ref{2.5}) which broadens the sound peaks in $S(k,\omega)$, giving them a line width that scales ``diffusively'' as $\Gamma(k)\sim k^2$.
This works fine in dimension $d \geq 3$  \cite{LL}, but fails in one dimension \cite{EHL}.  An example showing this anomaly is simulations of Fermi-Pasta-Ulam (FPU) chains which report
superdiffusive broadening of the sound peaks \cite{LLP1,LLP2,LLP3,dhar_review,hs2014}.  A  FPU chain consists of masses coupled  to their nearest neighbors through
anharmonic potentials. The discrete GPE has a similar structure, although instead of being anharmonic in the displacements, it is anharmonic in the local density.
To capture such anharmonic behavior, it has been proposed recently to use a nonlinear extension of fluctuating hydrodynamics
\cite{spohn}.  We will follow this strategy to obtain the hydrodynamic scaling of $S(k,\omega)$, which we then compare to exact Hamiltonian numerics.
The prescription of nonlinear fluctuating hydrodynamics \cite{spohn} consists of adding diffusion and noise matrices in Eq. \ref{2.4} giving,
\begin{equation}\label{2.6}
\partial_{t}\vec{u}+\partial_{x}\left[A\vec{u}+\frac{1}{2}\sum_{\alpha,\beta=1}^{2}\vec{H}_{\alpha,\beta}u_{\alpha} u_{\beta}-\partial_{x}\left(D\vec{u}\right)+B\vec{\xi}\right]=0
\end{equation}
where $D$ and $B$ are diffusion and noise matrices. Above, the vector $H_{\alpha,\beta}^{\gamma}= \partial_{u_{\alpha}}\partial_{u_{\beta}}j^{\gamma}$ with $\vec{j}=(\varrho v,\frac{1}{2}v^2)$.


Dropping the quadratic terms would correspond to linear fluctuating hydrodynamics and thus yield diffusive sound peaks.
For our application we are interested in the stationary, mean zero process governed by Eq. (\ref{2.6}), again  denoted by
$(\varrho(x,t), v(x,t))$. The equal time,
static correlations are expected to have short range correlations. Hence we define the susceptibilities
\begin{equation}\label{2.6a}
c_1 = \int dx (\langle \varrho(x,0)\varrho(0,0)\rangle - \rho_0^{\,2})\,, c_2 = \int dx \langle v(x,0)v(0,0)\rangle\
\end{equation}
where the cross terms vanish because $\varrho(x,t)$ and $v(x,t)$ have different parity.
The fluctuation dissipation relation is given by $DC+CD=BB^{\dagger}$  ($C$ is a diagonal matrix containing Eq. \ref{2.6a}).
  In addition, space-time stationarity enforces in general the relation $AC = CA^{\mathrm{T}}$, which implies $c_2 = \frac{c^2}{\rho_0^2} c_1$.
In Eq. (\ref{2.4}) the linear terms dominate and to obtain better insight to the solution one has to transform to
normal modes which have a definite propagation velocity. Thus one introduces a linear transformation in component space,
by setting
\begin{equation}\label{2.7}
\begin{pmatrix}
\phi_{-}\\ \phi_{+}
\end{pmatrix}
= R
\begin{pmatrix}
\varrho\\ v
\end{pmatrix}
\end{equation}
such that $R$ satisfies
$RAR^{-1} = \mathrm{diag}(-c,c)\,.$
In addition we require that the $\phi$-susceptibilities are normalized to one,
 which means
$R \,\mathrm{diag}(c_1,\frac{c^2}{\rho_0^2} c_1)R^{\mathrm{T}}= 1\,.$
Up to an overall sign, $R$ is uniquely determined and given by
\begin{equation}\label{2.10}
R =\frac{1}{c\sqrt{2 c_1}}\begin{pmatrix}
 -c & \rho_0\\ c &\rho_0
\end{pmatrix}\,.
\end{equation}
Then the equation for the normal modes (i.e, the left and right chiral sectors) reads
\begin{equation}\label{2.11}
\partial_{t}\phi_{\sigma}+\partial_{x}\big[\sigma c\phi_{\sigma}+\langle \vec{\phi,}G^{\sigma}\vec{\phi}\rangle
%
-\partial_x(D_\mathrm{rot}\phi)_{\sigma}+(B_\mathrm{rot}\xi)_{\sigma}\big]=0
\end{equation}
with $\sigma = \pm $ and ``{\it rot}'' indicating the matrices rotated by R matrix, $D_\mathrm{rot}=RDR$ and $B_\mathrm{rot}=RB$. The coupling matrix is given by,
 \begin{equation}\label{2.12}
G^{-} = \frac{c}{2\rho_{0}}\sqrt{\frac{c_{1}}{2}}
\begin{pmatrix}
3 & 1\\
1 & -1
\end{pmatrix}\,\, ,
G^{+} = \frac{c}{2\rho_{0}}\sqrt{\frac{c_{1}}{2}}
\begin{pmatrix}
-1 & 1\\
1 & 3
\end{pmatrix}\,.
\end{equation}
Since Eq. (\ref{2.11}) is nonlinear, it is still difficult to compute the covariance $\langle \phi_{\sigma}(x,t)
\phi_{\sigma'}(0,0)\rangle$ for the mean zero, stationary process. Although obvious, one central observation is that in leading order
the two peaks in $S(x,t)$ separate linearly in time. Hence, in the equation (\ref{2.11}) for $\phi_\sigma$, the terms $\phi_\sigma \phi_{-\sigma}$
and $(\phi_{-\sigma})^2$ turn out to be irrelevant \cite{hs2014} compared to $(\phi_\sigma)^2$. Albeit they may effectively renormalize the non-universal coefficients (in front of all terms), they do not impact the universal properties. Therefore, preserving universality we can decouple Eq. \ref{2.11} into two components giving,
\begin{equation}\label{2.13}
\partial_{t}\phi_{\sigma}+\partial_{x}\big[\sigma c\phi_{\sigma}+G_{\sigma\sigma}^{\sigma}\phi_\sigma^2
-\partial_x(D_\mathrm{rot}\phi)_{\sigma}+(B_\mathrm{rot}\xi)_{\sigma}\big]=0
\end{equation}

Eq.~\ref{2.13} is the stochastic Burgers equation (KPZ in ``height function" $h_{\sigma}$ where $h_{\sigma}=\partial_x \phi_{\sigma}$) and for it the exact scaling function is available,
\begin{equation}\label{2.14}
\langle \phi_\sigma(x,t)\phi_\sigma(0,0)\rangle = (\lambda t)^{-2/3}f_{\mathrm{KPZ}}\big[(\lambda t)^{-2/3}(x - \sigma c t)\big]
\end{equation}
valid for large $x,t$. $\lambda$ is a non-universal coefficient, which here is explicitly calculated to be
\begin{equation}
\label{lamkpz}
\lambda = 2 \sqrt{2} |G_{\sigma\sigma}^{\sigma}|
\end{equation}
One should keep in mind that the value of $\lambda$ derived above (\ref{lamkpz}) will get renormalized \cite{hs2014} due to the discarded non-linearities as explained above.  Note that $\lambda$ does not depend on $D_{\mathrm{rot}}$ or $B_{\mathrm{rot}}$.  This says that, while some
dissipation and noise is needed to maintain stationarity, the asymptotic form of the correlation is dominated by  $G_{\sigma\sigma}^{\sigma}$.
The definition of $f_{\mathrm{KPZ}}$ is somewhat indirect. One first computes
a family of probability densities $p_x(s)$ indexed by $x$, in essence defined by a Fredholm determinant which has to be  evaluated numerically \footnote{M. Prah\"{o}fer, Exact scaling functions for one-dimensional stationary KPZ growth.
http://www-m5.ma.tum.de/KPZ.}.
Then
$f_{\mathrm{KPZ}}(x) = \int ds p_x(s)s^2\,$.
$f_{\mathrm{KPZ}}$ has the following properties: $f_{\mathrm{KPZ}}\geq 0$, $\int dxf_{\mathrm{KPZ}}(x)=1$, $f_{\mathrm{KPZ}}(x)=f_{\mathrm{KPZ}}(-x)$, $\int dxf_{\mathrm{KPZ}}(x)x^2=0.510523$ and so on. $f_{\mathrm{KPZ}}$ looks like a Gaussian, but with a large $|x|$ decay instead as $\exp[-0.295|x|^{3}]$ \cite{ps04}. For a few discrete models there is in fact a proof
\cite{ps04,fs06}. For the stochastic Burgers equation there is a tricky replica computation yielding the same result \cite{Imamura2}.
In some molecular dynamics simulations, one records directly the correlation $S(x,t)$ \cite{PhysRevLett.111.230601} where $x$ is the space coordinate. But more conventionally,
as also relavant in this Letter, one studies the structure function, which is defined as the space-time Fourier transform of
$S(x,t)$.
We define $\hat{S}(k,t) =\int_{-\infty}^{\infty} \mathrm{e}^{- \mathrm{i} kx} S(x,t) dx $.
As argued before, the asymptotic scaling form is expected to be of the form
\begin{equation}\label{3}
\hat{S}(k,t) = \tfrac{1}{2}(\mathrm{e}^{\mathrm{i}kct} + \mathrm{e}^{-\mathrm{i}kct}) c_1 \hat{f}_{\mathrm{KPZ}}(k(\lambda  |t|)^{2/3})\,.
\end{equation}
The two sound peaks are symmetric reflections of each other. Considering only the right mover and setting $\omega_{k} = ck$, one concludes
\begin{eqnarray}\label{4}
&&\hspace{-20pt}\hat{S}(k,\omega+\omega_{\mathrm{k}}) = \int dt \mathrm{e}^{\mathrm{i}\omega t} \tfrac{1}{2} c_1 \hat{f}_{\mathrm{KPZ}}
(k(\lambda |t|)^{2/3})\\
&&\hspace{20pt}= \int dt \mathrm{e}^{\mathrm{i}(\omega/\lambda |k|^{3/2}) t} (\lambda |k|^{3/2})^{-1}
\tfrac{1}{2} c_1 \hat{f}_{\mathrm{KPZ}}(|t|^{2/3})\nonumber.
\end{eqnarray}
Thus defining
 \begin{equation}\label{5}
h(\omega) = \int dt \mathrm{e}^{\mathrm{i}\omega t}  \hat{f}_{\mathrm{KPZ}}(|t|^{2/3})\,,
\end{equation}
one arrives at
 \begin{equation}\label{6}
\hat{S}(k,\omega+\omega_{\mathrm{k}}) = \tfrac{1}{2} c_1 (\lambda |k|^{3/2})^{-1} h(\omega/ \lambda |k|^{3/2})\,.
\end{equation}
If  the maximum of $\hat{S}$ is normalized to 1, then the prefactor in (\ref{6}) is set to 1 and $h$ is replaced by $h/h(0)$.

\begin{figure}[ht]
\includegraphics[width=8.3cm]{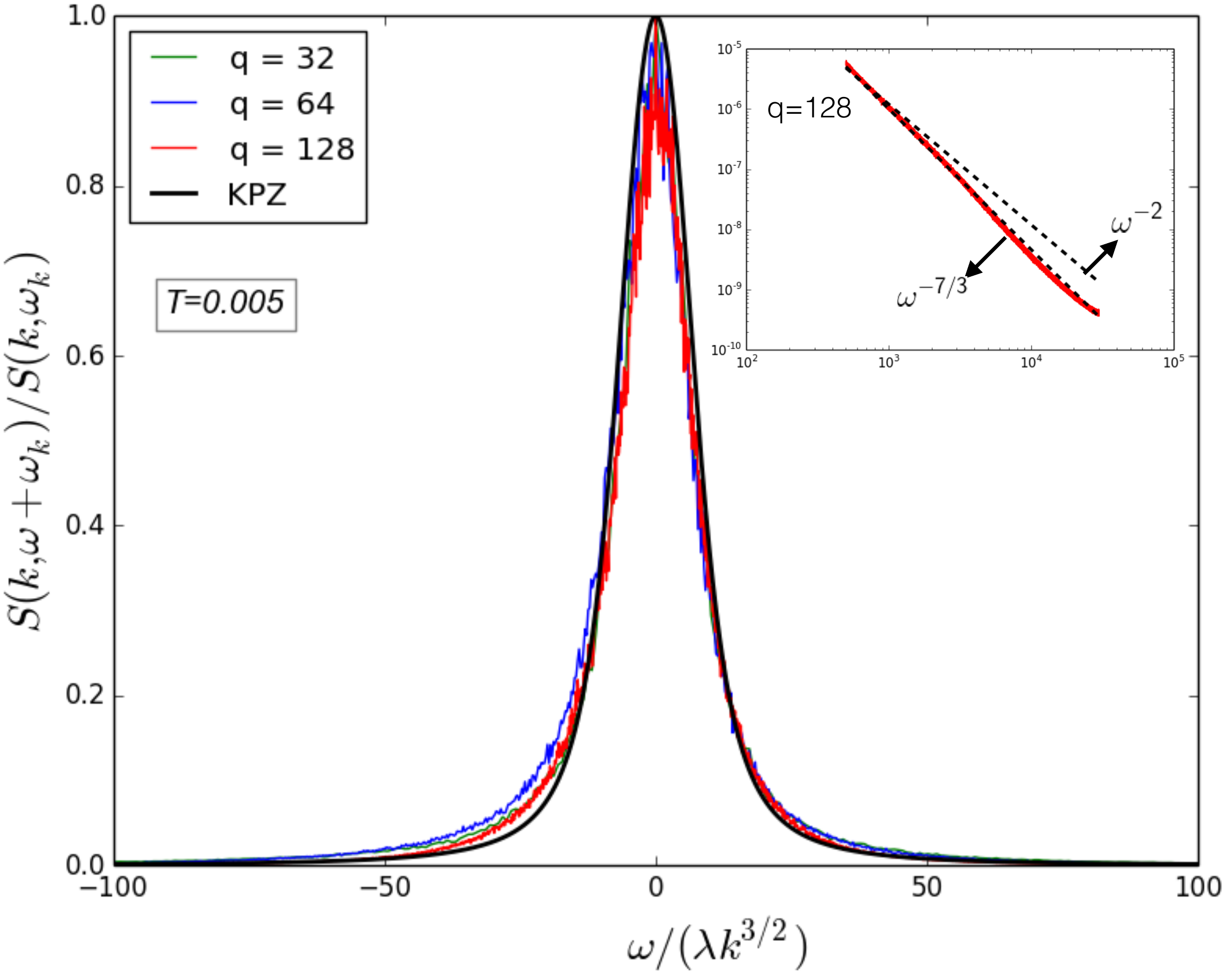}
\includegraphics[width=8.5cm]{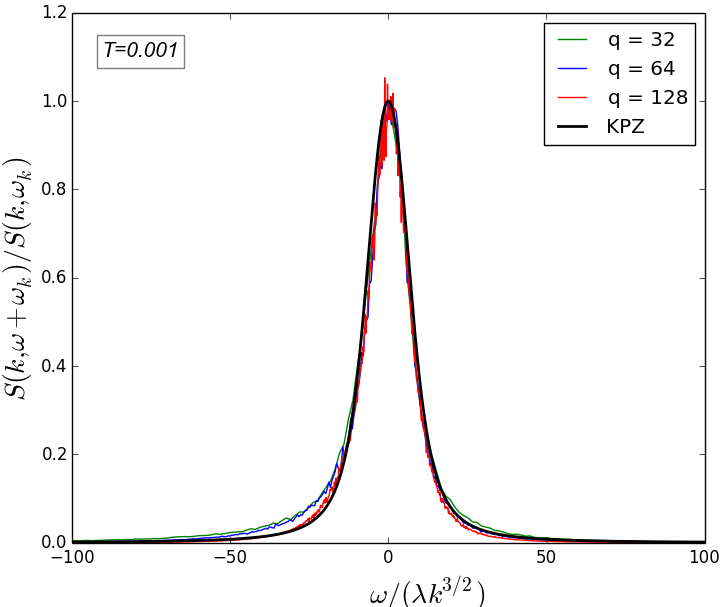}
\caption{(Top) Comparision between exact numerics of discrete GPE with nonlinear fluctuating hydrodynamics. Here the temperature $T=0.005$, $L= 5\times2^{14}$. The best fit to the scaling function $h(\omega)/h(0)$ is given by $\lambda_{opt} \sim 0.005$ (and theoretical $\lambda \sim 0.045$) with universal tail (shown by Log-Log inset) close to $\omega^{-\frac{7}{3}}$ at significantly large frequencies ~\cite{ps04}. (Bottom) Similar comparison for a different set of parameters. Here the temperature $T=0.001$, $L= 10\times2^{13}$. The best fit to the scaling function $h(\omega)/h(0)$ is given by $\lambda_{opt} \sim 0.0028$ (and theoretical $\lambda \sim 0.0041$). The wave vectors are given by $k=\frac{2\pi}{L} q$ where the values of the integers $q$ are given in the legends.}
\label{fig:NLShydro}
\end{figure}

One should cautiously note that the decoupling hypothesis is a subtle issue. The $\phi_\sigma(x,t)$ fields fluctuate without any
spatial decay. It is in fact, only the correlations that are peaked near $\pm c t$. However the decoupling of the components can be seen directly on the level of mode-coupling in the one-loop approximation. As supported by numerical solutions \cite{PhysRevLett.111.230601}, it is safe to use the diagonal approximation
\begin{equation}\label{2.16}
\langle \phi_{\sigma}(x,t)\phi_{\sigma'}(0,0)\rangle = \delta_{\sigma\sigma'}  f_\sigma (x,t)   \,.
\end{equation}
Of course $f_\sigma (x,t) = f_{-\sigma} (-x,t)$. In one-loop one has ($\nu$ is the phenomenologically added dissipation)
\begin{eqnarray}\label{2.17}
\partial_t   f_\sigma (x,t) &=& (-\sigma c \partial_x + \tfrac{1}{2}\nu \partial_x^2) f_\sigma (x,t)\nonumber\\&+& \int_0^tds \int dyf_\sigma (x-y,s) \partial_y^2M(y,s)
\end{eqnarray}
 with the memory kernel
\begin{equation}\label{2.18}
M(x,t) = 2 \sum_{\sigma\sigma'= \pm}(G_{\sigma\sigma'}^{\sigma})^2 f_\sigma (x,t)f_{\sigma'} (x,t)\,.
\end{equation}
The terms with $\sigma \neq \sigma'$ have a very small overlap. But the diagonal terms proportional to
$(G_{\sigma\sigma}^{\sigma})^2$ do contribute to the long time behavior. By explicit computation one checks that the self-interaction
term dominates the mutual one. Eq. (\ref{2.17}) can be studied numerically by an iteration scheme. The asymptotic shape of the sound
peak is, of course, not the true scaling function $f_{\mathrm{KPZ}}$, but so with a relative error of about $4\%$ \cite{PhysRevLett.111.230601}. It is of utmost importance to have such a deterministic expression (\ref{2.17}) for the correlators of 1D Bose gas that captures the physics beyond a conventional Luttinger Liquid. This computationally advantageous method (Eq. \ref{2.17}) along with our derived Eq. \ref{lamkpz} and the above established stochastic nonlinear field theoretic description of 1D Bose gas needs to be benchmarked against brute-force Hamiltonian numerics of the underlying GPE Hamiltonian.

\textit{Hamiltonian numerics of discrete-GPE: }
We now go to the discrete version of above time-dependent GPE (Eq. \ref{tGPE}) that now governs the dynamics of a complex-valued $\psi(n,t)$, with integer
$n= 1,...,N$ and periodic boundary conditions. Discretization is achieved by substituting $x \rightarrow na$ where $a$ is the lattice spacing and $Na$ is the system size $L$.
The discrete version of time-dependent GPE reads,
\begin{equation}\label{2.2}
i\frac{d}{dt} \psi(n,t) = \mathcal{IF}\bigg[\frac{k_q^2}{2m} \tilde \psi(k_q,t)\bigg]
+ g  |\psi(n,t)|^2\psi(n,t)
\end{equation}
where $k_{q}=\frac{2\pi}{Na}q$ for integer $q=\big(-\frac{N}{2}+1\big),...,\frac{N}{2}$ and $\mathcal{IF}$ denotes our inverse-Fourier transform, $\mathcal{IF}\{G(k_q)\}=\frac{1}{a}\frac{1}{N} \sum_{n=1}^{N} G(k_q)e^{-\frac{2\pi i}{N}nq}$ (slightly unconventional due to explicit presence of lattice spacing $a$).
The local energy and the local number density,
$|\psi(n)|^2$, are conserved. According to standard classifications, Eq. (\ref{2.2}) is listed as not integrable  \cite{APT}.
Hence one would expect that $H$ and $\mathsf{N} = a\sum_{n=1}^N|\psi(n)|^2$ are the only conserved fields and that the set  of equilibrium states is of the form $Z^{-1} e^{-\beta(H -\mu \mathsf{N})}$, $\beta >0$, $\mu \in \mathbb{R}$, in the limit of large $N$.
Therefore, the above discretization scheme for the integrable continuum GPE breaks the underlying integrability.  In fact, we find that, in order to make connection to fluctuating hydrodynamics and subsequently KPZ, we require broken integrability and the resulting chaos.



In this section, we describe the Hamiltonian exact numerics  \cite{kulkarni_lamacraft} starting from Eq. \ref{2.2}. The time evolution
is obtained by the well-known leap frog splitting technique where the system is evolved alternatively (setting $g=m=1$, and always choosing $\rho_0=1$) by kinetic, $\tilde{\psi}(k_q,t) \rightarrow e^{-ik_q^{2}\tau/2}\tilde{\psi}(k_q,t)$, and potential, $\psi (n,t) \rightarrow e^{-i\tau|\psi(n,t)|^{2}} \psi (n,t)$, terms in sequence $\mathcal{V}_{\frac{\tau}{2}}\cdot\mathcal{T_{\tau}}\cdot\mathcal{V}_{\frac{\tau}{2}}$, with time step $\tau$.

In the simulation \cite{kulkarni_lamacraft} we measure the structure function $\tilde{S}(k_q,\omega)$:
At each time step we
obtain the time evolved density $\rho(n,t)=|\psi(n,t)|^2$, which we then space-time Fourier transform to $\tilde\rho(k_q,\omega)$.
Then the dynamical structure factor is the ensemble average:
$\tilde S(k_q,\omega)=\langle|\tilde\rho(k_q,\omega)|^2\rangle$.  These results are expected to depend only on the total energy and
particle number of the initial condition, since the dynamics are chaotic and we expect ergodic.
The chaos for our parameters $a=5$ and $\tau=2$ has been confirmed by observing positive Lyapunov exponents.

 For the random initial conditions we assume that the Fourier coefficients $\varrho_{k},\theta_{k}$ are independent Gaussian random variables with mean 0 and covariance given by
\begin{eqnarray}
\left\langle \left|\varrho_{k_{q}}^{2}\right|\right\rangle = \frac{\rho_0}{2L}\frac{\alpha_{k_{q}} T}{ \xi_{k_q}}, \qquad\left\langle \left|\theta_{k_{q}}^{2}\right|\right\rangle = \frac{1}{2\rho_0 L}\frac{T}{\alpha_{k_{q}} \xi_{k_q}}
\end{eqnarray}
%
where $\xi_{k_q}=\sqrt{\frac{k_q^{2}}{2}\left(\frac{k_q^{2}}{2}+2\rho_0\right)}$ and  $\alpha_{k_q}=\sqrt{\frac{k_q^{2}}{\frac{k_q^{2}}{4}+\rho_0}}$.



In Ref. \cite{kulkarni_lamacraft} the low temperature dynamical structure factor was simulated numerically and the KPZ scaling exponent was observed,
i.e. phonon line width $\Gamma_k \sim |k|^z$ with $z=1.510\pm 0.018$.
Here we provide a more quantitative comparison with the full scaling function (Fig. \ref{fig:NLShydro}).
We also have fully outlined the mapping to nonlinear fluctuating hydrodynamics, which tells us that the structure factor should be of the form shown
in Eq. (\ref{6}).  This means that the structure factor we obtain from our simulations must scale with the KPZ scaling exponent and
scaling function in the hydrodynamic limit.

In Fig. \ref{fig:NLShydro} we show the remarkable quantitative agreement between exact Hamiltonian numerics and the expectations of a nonlinear hydrodynamic theory with fluctuations. 
Our results are in a regime where the system is not near integrability, due to the large lattice spacing $a$.  We have checked that on approaching integrability by reducing $a$ we find strong deviations from KPZ, both in terms of scaling form and exponent, as we enter the regime of crossover between integrability and KPZ scaling.  The discrepancy between the optimally chosen value of $\lambda$ ($\lambda_{opt}$) and the one expected from KPZ correspondence (Eq. \ref{lamkpz}) probably arises due the fact that higher-order nonlinearities and the different chiral sectors effectively renormalize the first relevant nonlinearity of the specific chiral sector under consideration. Such a disagreement has also been seen recently in case of the FPU problem~\footnote{M. Straka, KPZ scaling in the one-dimensional FPU $\alpha - \beta$ model. Master’s thesis, University of Florence, Italy (2013) \label{f1}}. One does require an effective renormalization scheme to make more precise connections between $\lambda$ and $\lambda_{opt}$.

In conclusion, we demonstrate a strong connection between the statistical mechanics of a discrete NLS/GPE and a nonlinear hydrodynamic theory with fluctuations. This was done by first formulating the GPE in terms of hydrodynamic variables (conjugate classical fields) and then adapting a recent procedure in formulating a fluctuating version of the nonlinear hydrodynamic theory~\cite{spohn}. In our case, the resulting theory is shown to be of the KPZ universality class. This immediately enables us to use the rich physics of KPZ class and a well-established one-loop approximation to make predictions for GPE. This was then benchmarked by exact numerics of the molecular dynamics type. Given the wide range of phenomena described by these equations, our results have implications in fields ranging from cold gases to nonlinear optics. Moreover, extending this mapping to coupled Nonlinear Schr\"{o}dinger equations (also an experimentally realized situation in both cold atoms~\cite{jorg,zhaihui} and nonlinear optics~\cite{kang}) is shown to give a whole zoo of interesting dynamical critical phenomenon arising due to coupled stochastic differential equations~\footnote{M. Kulkarni, D. A. Huse,  H. Spohn, 2015 (Unpublished)}. \\
\\
We thank A. Lamacraft and A. Dhar for enlightening discussions. We acknowledge the hospitality of the School of Mathematics at the Institute for Advanced Study, Princeton during the program on ``Non-equilibrium Dynamics and Random Matrices'' where several discussions took place. M. K. thanks the hospitality of the International Centre for Theoretical Sciences, Bengaluru and Tata Institute of Fundamental Research, Mumbai where there were many fruitful discussions.

\bibliographystyle{apsrev}
\bibliography{GPE_KPZ}

\end{document}